\documentclass[prb,twocolumn,showpacs,aps]{revtex4}
\usepackage{graphicx,graphics,color,epsfig}
\usepackage{bm}
\usepackage{amsmath}
\usepackage{amssymb}
\usepackage{epstopdf}

\makeatletter

\newcommand{\Rmnum}[1]{\expandafter\@slowromancap\romannumeral #1@}
\makeatother

\begin{document}

\title{Mixed-State Effect on the Low-Energy Spin Dynamics in Optimally-doped Iron Pnictide Superconductors}

\author{Yi Gao,$^{1}$ Tao Zhou,$^{2}$ Huaixiang Huang,$^{3}$ C. S. Ting,$^{4}$ and Peiqing Tong$^{1}$}
\affiliation{$^{1}$Department of Physics and Institute of Theoretical Physics,
Nanjing Normal University, Nanjing, Jiangsu, 210023, China\\
$^{2}$College of Science, Nanjing University of Aeronautics and
Astronautics, Nanjing, Jiangsu, 210016, China\\
$^{3}$Department of Physics, Shanghai University, Shanghai, 200444, China\\
$^{4}$Department of Physics and Texas Center for Superconductivity, University of Houston, Houston, Texas, 77204, USA}

\begin{abstract}

Based on a phenomenological model with $s_{\pm}$ or $s$-wave pairing symmetry, the mixed-state effect on the low-energy spin dynamics in optimally-doped iron pnictide superconductors is studied by solving Bogoliubov-de Gennes equations. Our results of the spin susceptibility at $\mathbf{q}=\mathbf{Q}$ in the normal, superconducting and mixed states agree qualitatively with recent neutron scattering experiments. We also propose that the field-induced intensity change shows different behaviors between the $s_{\pm}$ and $s$-wave symmetries in both momentum and real space, thus it can be used to distinguish these two pairing symmetries.
\end{abstract}


\maketitle

\emph{Introduction}.---The superconducting (SC) pairing mechanism and symmetry in iron pnictide superconductors have long been debated even since their discovery in 2008.~\cite{I1} Theoretically, one commonly believed idea is that the Cooper pairs are formed via spin fluctuations, leading to the $s_{\pm}$ pairing symmetry where the paring order parameter (OP) changes sign between the electron- and hole-pockets.~\cite{I20,I20_4,I20_5,I20_1,I20_2,I20_3} Later, orbital-fluctuation-mediated pairing was also proposed and the symmetry in this case is $s$-wave without sign reversal.~\cite{s++,s++_1} Experimentally, the evidences for the paring symmetry ($s_{\pm}$ or $s$) are not definite either. For example, in optimally-doped Ba$_{0.6}$K$_{0.4}$Fe$_{2}$As$_{2}$ and BaFe$_{1.85}$Co$_{0.15}$As$_{2}$, the SC gaps measured by angle-resolved photoemission spectroscopy (ARPES) are almost isotropic on both the electron- and hole-pockets.~\cite{I21-10,I21-10_1} However, due to the insensitivity of ARPES to the sign of the OP, the paring symmetry can be either $s_{\pm}$ or $s$-wave. Neutron scattering (NS) experiments observed a resonance at $\textbf{Q}=(\pm\pi,0)$ and $(0,\pm\pi)$ (in the 1Fe/cell Brillouin zone) in the SC state,~\cite{ns,ns_2,ns_3,ns_4,ns_5,ns_1} which was initially interpreted as the evidence for the $s_{\pm}$ symmetry.~\cite{nstheory,nstheory_1,nstheory_2,nstheory_3} Later it was found that the resonance can also emerge if the symmetry is $s$-wave.~\cite{nstheory2,nstheory2_1} Furthermore, the local density of states probed by scanning tunneling microscopy (STM) with or without the magnetic field~\cite{stm} can be reproduced by assuming either the $s_{\pm}$ or $s$-wave symmetry.~\cite{gao2}

Recently Hanaguri \emph{et al.} used STM to image the quasiparticle interference (QPI) and stated that the QPI in the presence of a magnetic field can be used to distinguish the $s_{\pm}$ and $s$-wave symmetries,~\cite{hanaguri} which has been verified by our previous work.~\cite{gao3} In addition, some NS experiments studied the effect of a magnetic field on the low-energy spin dynamics.~\cite{m1,m2,m3,m4,m5} Specifically in Ref.~\onlinecite{pcdai}, a perpendicular magnetic field applied in optimally-doped BaFe$_{1.9}$Ni$_{0.1}$As$_{2}$ reduces the intensity and energy of the resonance while broadens its width. A natural question arises: Is the evolution of the resonance with the magnetic field consistent with the $s_{\pm}$ or $s$-wave scenario and are there differences between them as in Ref.~\onlinecite{hanaguri}? In this paper we address this issue and propose a method to test it. The mix-state effect on the spin dynamics is studied by solving the Bogoliubov-de Gennes (BdG) equations in the presence of a perpendicular magnetic field and then conducting the real-space random-phase approximation (RPA) to calculate the spin susceptibility $\chi$. Previously, for the iron pnictides without the magnetic field, $\chi$ has been calculated in momentum space.~\cite{nstheory,nstheory_1,nstheory_2,nstheory_3,nstheory2,nstheory2_1} However in the presence of a magnetic field, the translational symmetry is broken and $\chi$ has to be calculated in real space which requires much more computational time and resources. Therefore a theoretical investigation of the mixed-state effect on the spin dynamics in the iron pnictides is still lacking and needs to be done urgently. By studying this, we find that while the spin excitation spectra in the $s_{\pm}$ and $s$-wave cases are similar to each other no matter the magnetic field is present or not, there are clear differences in the field-induced intensity change which can be used to distinguish the two pairing symmetries.

\emph{Method}.---We adopt an effective two-orbital model on a
two-dimensional $N\times N$ lattice which captures the basic Fermi surface structures of the iron pnictides,~\cite{M55} with a phenomenological form for
the intraorbital pairing terms. The Hamiltonian can be written as~\cite{jiang,zhou,H}
\begin{eqnarray}
\label{h0}
H_{0}=&-&\sum_{ij,\alpha\beta,\nu}t^{'}_{ij,\alpha\beta}c^{\dag}_{i\alpha\nu}c_{j\beta\nu}+\sum_{j\beta\nu}\big{[}-\mu+U\langle n_{j\beta\bar{\nu}}\rangle\nonumber\\
&+&(U-2J_{H})\langle n_{j\bar{\beta}\bar{\nu}}\rangle+(U-3J_{H})\langle n_{j\bar{\beta}\nu}\rangle\big{]}n_{j\beta\nu}\nonumber\\
&+&\sum_{ij,\alpha\beta}(\Delta_{ij,\alpha\beta}c^{\dag}_{i\alpha\uparrow}c^{\dag}_{j\beta\downarrow}+H.c.).
\end{eqnarray}
Here $i,j$ are the site indices, $\alpha,\beta=1,2$ are the orbital indices, $\nu$ represents the spin, $\mu$ is the
chemical potential and $n_{j\beta\nu}=c^{\dag}_{j\beta\nu}c_{j\beta\nu}$ is the number operator. $U$ and $J_{H}$ are the onsite intraorbital Hubbard repulsion and Hund's coupling, respectively. Here we have the interorbital Coulomb interaction $U^{\prime}=U-2J_{H}$ according to symmetry.~\cite{H_1} $\Delta_{ij,\alpha\beta}=\frac{V_{ij}\delta_{\alpha\beta}}{2}(\langle c_{j\beta\downarrow}c_{i\alpha\uparrow}\rangle-\langle c_{j\beta\uparrow}c_{i\alpha\downarrow}\rangle)$ is the intraorbital spin-singlet bond OP, where $V_{ij}$ is the next-nearest-neighbor [$i=j\pm(\hat{x}\pm\hat{y})$] or onsite [$i=j$] attraction we choose to achieve the $s_{\pm}$ or $s$-wave symmetry, respectively. In the presence of a magnetic field $B$ perpendicular
to the plane, the hopping integral is
$t^{'}_{ij,\alpha\beta}=t_{ij,\alpha\beta}$exp$[i\frac{\pi}{\Phi_{0}}\int_{j}^{i}\mathbf{A}(\mathbf{r})\cdot
d\mathbf{r}]$, where $\Phi_{0}=hc/2e$ is the SC flux quantum, and
$\mathbf{A}=(-By,0,0)$ is the vector potential in the Landau gauge.
Following Ref.~\onlinecite{M55},
\begin{equation}
\label{hopping}
t_{ij,\alpha\beta}=\begin{cases}
t_{1}&\text{$\alpha=\beta,i=j\pm\hat{x}(\hat{y})$},\\
\frac{1+(-1)^{j}}{2}t_{2}+\frac{1-(-1)^{j}}{2}t_{3}&\text{$\alpha=\beta,i=j\pm(\hat{x}+\hat{y})$},\\
\frac{1+(-1)^{j}}{2}t_{3}+\frac{1-(-1)^{j}}{2}t_{2}&\text{$\alpha=\beta,i=j\pm(\hat{x}-\hat{y})$},\\
t_{4}&\text{$\alpha\neq\beta,i=j\pm(\hat{x}\pm\hat{y})$},\\
0&\text{otherwise}.
\end{cases}
\end{equation}

Equation (\ref{h0}) can be diagonalized by solving
the BdG equations $H=C^{\dag}MC$, where
$C^{\dag}=(\cdots,c^{\dag}_{j1\uparrow},c_{j1\downarrow},c^{\dag}_{j2\uparrow},c_{j2\downarrow},\cdots)$,
subject to the self-consistency conditions:
$\langle n_{j\beta\uparrow}\rangle=\sum_{k=1}^{L}|Q_{m-1k}|^{2}f(E_{k})$,
$\langle n_{j\beta\downarrow}\rangle=1-\sum_{k}|Q_{mk}|^{2}f(E_{k})$ and
$\Delta_{ij,\alpha\beta}=\frac{V_{ij}}{2}\delta_{\alpha\beta}\sum_{k}(Q^{*}_{mk}Q_{nk}+Q^{*}_{n+1k}Q_{m-1k})f(E_{k})$.
Here $L=4N^{2}$, $m=4(j_{y}+Nj_{x})+2\beta$, $n=4(i_{y}+Ni_{x})+2\alpha-1$
and $Q$ is a unitary matrix that satisfies
$(Q^{\dag}MQ)_{kl}=\delta_{kl}E_{k}$. Here we used $i=(i_{x},i_{y})$
and $j=(j_{x},j_{y})$, with $i_{x},j_{x},i_{y},j_{y}=0,1,\ldots,N-1$. $\mu$ is
determined by the doping concentration $x$ through
$\frac{1}{N^{2}}\sum_{j\beta\nu}n_{j\beta\nu}=2+x$.

The bare spin susceptibility is~\cite{x}
\begin{eqnarray}
\label{x0}
\chi_{^{\delta\alpha}_{\gamma\beta}}^{-+0}(i,j,\omega)&=&-\langle\langle c^{\dag}_{i\delta\downarrow}c_{i\alpha\uparrow}|c^{\dag}_{j\gamma\uparrow}c_{j\beta\downarrow}\rangle\rangle_{\omega+i\eta}\nonumber\\
&=&\sum_{k,l=1}^{L}Q_{pk}Q_{nl}(Q^{*}_{ol}Q^{*}_{mk}-Q^{*}_{ok}Q^{*}_{ml})\nonumber\\
&&\frac{f(E_{k})+f(E_{l})-1}{\omega-E_{k}-E_{l}+i\eta},
\end{eqnarray}
where $o=4(j_{y}+Nj_{x})+2\gamma-1$ and $p=4(i_{y}+Ni_{x})+2\delta$. Including interactions within RPA, the full susceptibility can be calculated through
\begin{eqnarray}
\label{x}
\chi_{^{\delta\alpha}_{\gamma\beta}}(i,j,\omega)&=&\chi_{^{\delta\alpha}_{\gamma\beta}}^{0}(i,j,\omega)+\sum_{v}\sum_{rstu}\chi_{^{\delta\alpha}_{rt}}^{0}(i,v,\omega)\nonumber\\
&&U^{I}_{^{rt}_{su}}(v,v)\chi_{^{su}_{\gamma\beta}}(v,j,\omega).
\end{eqnarray}
Here we omitted the superscript $-+$. The interaction vertex $U^{I}_{^{rt}_{su}}(i,j)$ is nonzero only when $i=j$ and satisfies
\begin{eqnarray}
\label{us}
U^{I}_{^{rr}_{rr}}=U,U^{I}_{^{rt}_{rt}}=J_{H},U^{I}_{^{rt}_{tr}}=U-2J_{H},U^{I}_{^{rr}_{tt}}=J_{H},
\end{eqnarray}
where $r\neq t$. The experimentally measured susceptibility is proportional to $\chi^{''}(\mathbf{q},\omega)$, where
\begin{eqnarray}
\label{xq}
\chi^{''}(\mathbf{q},\omega)&=&\frac{1}{N^{2}}\sum_{ij,\alpha\gamma}Im\chi_{^{\alpha\alpha}_{\gamma\gamma}}(i,j,\omega)\cos\mathbf{q}\cdot(i-j)\nonumber\\
&=&\sum_{i-j}f(i-j,\omega)\cos\mathbf{q}\cdot(i-j),
\end{eqnarray}
with $f(i-j,\omega)=\frac{1}{N^{2}}\sum_{j,\alpha\gamma}Im\chi_{^{\alpha\alpha}_{\gamma\gamma}}(i-j,j,\omega)$ being the average real-space spin-spin correlation function.

The parameters are chosen as $t_{1-4}=1,0.4,-2,0.04$. $N=32$ and magnetic unit cells are introduced where
each unit cell accommodates two SC flux quanta. $V_{ij}$
[$i=j\pm(\hat{x}\pm\hat{y})$] and $V_{ii}$ are chosen to be $-2$ and $-3.2$,
respectively. We fix $x=0.1$, corresponding to the optimally-doped BaFe$_{1.9}$Ni$_{0.1}$As$_{2}$, which enables us to compare the results directly with those in Ref.~\onlinecite{pcdai}.
The temperature $T$ is set to be $0$ and $0.7\Delta_{0}$. Here $\Delta_{0}$ is the SC OP we solved at $(B=0,T=0)$. At $T=0.7\Delta_{0}$, the SC OP is $0$, indicating that the system is in the normal state. In the following we calculate $\chi^{''}(\mathbf{q},\omega)$ at $(B=0,T=0.7\Delta_{0})$, $(B=0,T=0)$ and $(B\neq0,T=0)$, corresponding to the normal, SC and mixed states, respectively.

\begin{figure}
\includegraphics[width=1\linewidth]{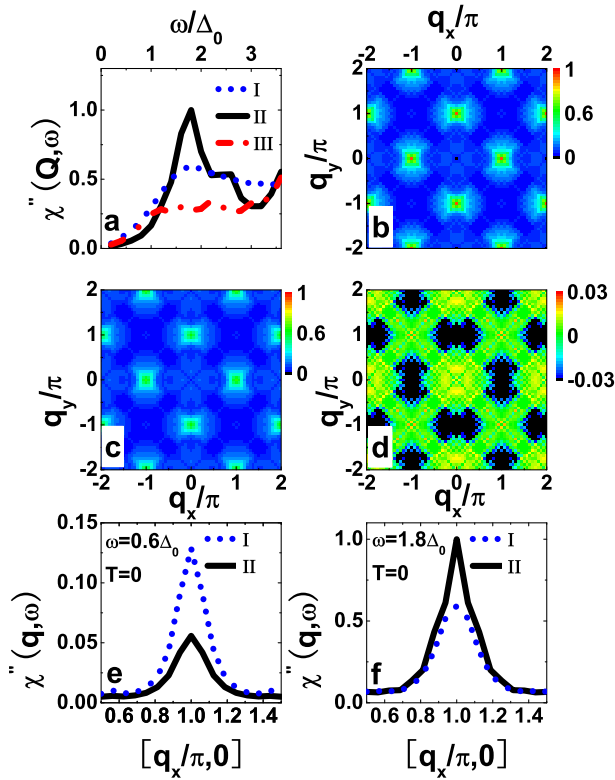}
 \caption{\label{fig1} (color online) The $s_{\pm}$ case. (a) Normalized $\chi^{''}(\mathbf{Q},\omega)$ as a function of the reduced energy $\frac{\omega}{\Delta_{0}}$.  (b) and (c) show $\chi^{''}(\mathbf{q},\omega_{res})$ in the SC and mixed states, respectively. (d) The field-induced intensity change $\Delta\chi^{''}(\mathbf{q},\omega_{res})$. (e) and (f) show $\chi^{''}(\mathbf{q},\omega)$ along the $(\mathbf{q}_{x},0)$ direction at $T=0$, for $\omega=0.6\Delta_{0}$ and $\omega=\omega_{res}$, respectively. In (a), (e) and (f), \Rmnum{1} (blue dot), \Rmnum{2} (black solid) and \Rmnum{3} (red dash dot) represent the mixed, SC and normal states, respectively.}
\end{figure}

\emph{Results}.---For the $s_{\pm}$ pairing, we choose $U$ and $J_{H}$ to be $2$ and $0.2U$, respectively. With this choice of parameters, the system stays paramagnetic no matter it is in the normal, SC or mixed state.  Meanwhile $\chi^{''}(\mathbf{Q},\omega)$ in the SC state is enhanced around $\omega\approx2\Delta_{0}$ as compared to its normal-state counterpart. Both the disappearance of the magnetic order and the enhancement of spin susceptibility around $2\Delta_{0}$ are typical characteristics of the optimally-doped iron pnictides~\cite{ns,ns_2,ns_3,ns_4,ns_5,ns_1,pd_1,pd} and the results shown below remain valid as long as $U$ and $J_{H}$ are in a reasonable regime.  In this case $\chi^{''}(\mathbf{q},\omega)$ in the SC state strongly peaks at $\mathbf{q}=\mathbf{Q}$ and $\omega=1.8\Delta_{0}$ ($\omega_{res}$), so in the following we normalize $\chi^{''}(\mathbf{q},\omega)$ by the value of $\chi^{''}(\mathbf{Q},\omega_{res})$ in the SC state. The constant-$\mathbf{q}$ scan in Fig. \ref{fig1}(a) shows that, the normal-state $\chi^{''}(\mathbf{Q},\omega)$ (red dash dot) exhibits clear gapless continuum of excitation. Upon entering the SC state, a spin gap gradually opens below $\omega\approx\Delta_{0}$ where the value of $\chi^{''}(\mathbf{Q},\omega)$ becomes smaller than that of the normal state and the low-energy spectral weight is transferred into a resonance at $\omega=\omega_{res}$ (black solid), consistent with previous momentum-space calculations without the magnetic field,~\cite{nstheory,nstheory_1,nstheory_2,nstheory_3} thus justifying the validity of our method. While in the normal-state, $\chi^{''}(\mathbf{q},\omega)$ is almost unaffected by the magnetic field, in the mixed state, vortices emerge and the intensity of the resonance in $\chi^{''}(\mathbf{Q},\omega)$ is clearly suppressed (blue dot) compared to that in the SC state. Furthermore the resonance shifts to a lower energy ($\omega=1.6\Delta_{0}$) and its width becomes broader, together with an intensity gain below $\omega\approx1.4\Delta_{0}$. The behavior of $\chi^{''}(\mathbf{Q},\omega)$ is thus qualitatively consistent with the experimental observation (see Fig. 1 in Ref.~\onlinecite{pcdai}). Figure \ref{fig1}(b) plots $\chi^{''}(\mathbf{q},\omega_{res})$ in the SC state. It strongly peaks at $\mathbf{q}=\mathbf{Q}$ where the spectrum is two-fold symmetric and shows a slight transverse-elongation, similar to that measured in Ref.~\onlinecite{pcdai_1}. In the mixed state, the spin excitation remains commensurate and $\chi^{''}(\mathbf{q},\omega_{res})$ does not change much except for that at $\mathbf{q}=\mathbf{Q}$, where the intensity is strongly reduced [Fig. \ref{fig1}(c)]. Figure \ref{fig1}(d) shows the field-induced intensity change defined as
$\Delta\chi^{''}(\mathbf{q},\omega_{res})=\chi^{''}(\mathbf{q},\omega_{res})|_{mixed-state}-\chi^{''}(\mathbf{q},\omega_{res})|_{SC-state}$.
We can see that the intensity is lowered around $\mathbf{q}=\mathbf{Q}$ and the spectrum also elongates transversely. At $T=0$ and $\omega=0.6\Delta_{0}$ (near the spin-gap energy), constant-$\omega$ scan along the $(\mathbf{q}_{x},0)$ direction [Fig. \ref{fig1}(e)] suggests that the scattering in the SC state is very weak, with a small peak around $\mathbf{q}=\mathbf{Q}$, consistent with the presence of a spin gap. The magnetic field induces scattering in the spin gap and enhances the small peak around $\mathbf{q}=\mathbf{Q}$ since there are in-gap Andreev bound states emerging close to the vortex centers.~\cite{H} In contrast, the imposition of the magnetic field at $T=0$ partially suppresses the resonance intensity at $\omega=\omega_{res}$ [Fig. \ref{fig1}(f)].

\begin{figure}
\includegraphics[width=1\linewidth]{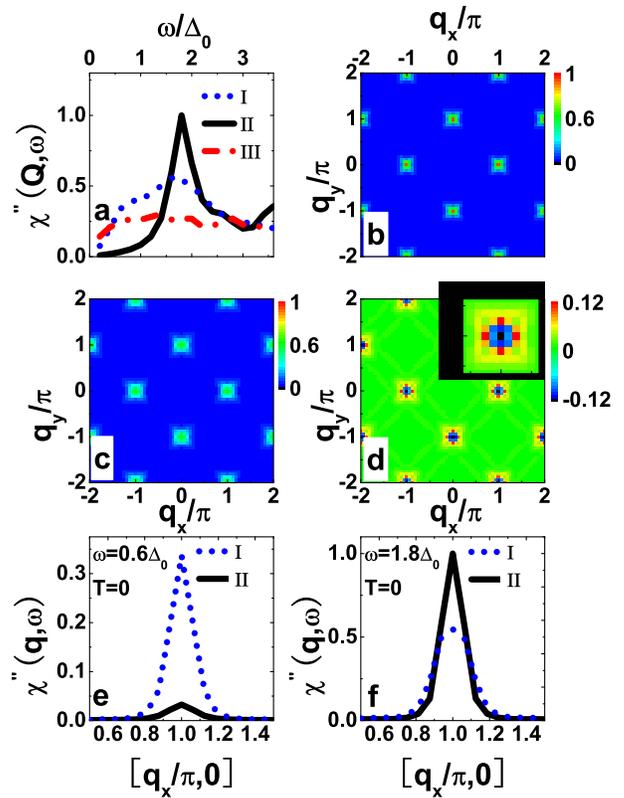}
 \caption{\label{fig2} (color online) The $s$-wave case. (a) to (f) are the same as Figs. \ref{fig1}(a) to \ref{fig1}(f), respectively. The inset in (d) shows the spectrum close to $\mathbf{q}=\mathbf{Q}$.}
\end{figure}

\begin{figure}
\includegraphics[width=1\linewidth]{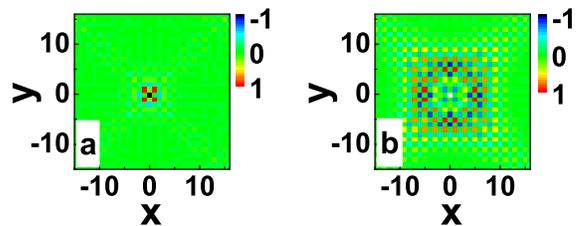}
 \caption{\label{fig3} (color online) $\Delta f(i-j,\omega_{res})$ for the $s_{\pm}$ (a) and $s$-wave (b) cases.}
\end{figure}

In the $s$-wave case, we set $U$ and $J_{H}$ to be $4.1$ and $0.5U$, respectively, for the same reason as mentioned above. If we still choose $U=2$ and $J_{H}=0.2U$, no resonance will appear below $\omega\approx5\Delta_{0}$ in the SC state. On the other hand, if we choose $U=4.1$ and $J_{H}=0.5U$ for the $s_{\pm}$ pairing, magnetic order will appear in the SC state and coexist with superconductivity. However for the same $U$ and $J_{H}$ in the $s$-wave case, no magnetic order exists in the SC state since it cannot coexist with $s$-wave superconductivity~\cite{fernandes} and this has been verified by our numerical calculation. Therefore we choose different $U$ and $J_{H}$ for the two pairing symmetries. The behavior of $\chi^{''}(\mathbf{Q},\omega)$ [Fig. \ref{fig2}(a)] shares some similarities with that in the $s_{\pm}$ case. The excitation is gapless in the normal state while a spin gap below $\omega\approx1.4\Delta_{0}$ and a resonance at $\omega=\omega_{res}$ show up in the SC state. Upon applying the magnetic field, the intensity of the resonance is suppressed and its energy is lowered to $\omega=1.6\Delta_{0}$. Similarly the resonance becomes broad and below $\omega=1.6\Delta_{0}$, the intensity increases due to the emergence of in-gap Andreev bound states near the vortex cores.~\cite{gao2} $\chi^{''}(\mathbf{q},\omega_{res})$ in the SC state [Fig. \ref{fig2}(b)] also peaks at $\mathbf{q}=\mathbf{Q}$, with four-fold symmetry in contrast to the two-fold one in the $s_{\pm}$ case. In the mixed state, the peak in $\chi^{''}(\mathbf{q},\omega_{res})$ at $\mathbf{q}=\mathbf{Q}$ is suppressed and the excitation area enlarges slightly [Fig. \ref{fig2}(c)]. The field-induced intensity change in Fig. \ref{fig2}(d) shows that, close to $\mathbf{q}=\mathbf{Q}$, the intensity is suppressed by the application of the magnetic field. However, further away from $\mathbf{q}=\mathbf{Q}$, there are intensity gains in the red and yellow areas. This behavior is different from that in the $s_{\pm}$ case where there are no intensity-gain areas around $\mathbf{q}=\mathbf{Q}$ as shown in Fig. \ref{fig1}(d). Thus the distinct characteristics in the field-induced intensity change can be used to distinguish the $s_{\pm}$ and $s$-wave symmetries. In the spin gap, the constant-$\omega$ scan along the $(\mathbf{q}_{x},0)$ direction at $T=0$ and $\omega=0.6\Delta_{0}$ [Fig. \ref{fig2}(e)] is qualitatively the same as that for the $s_{\pm}$ pairing, where the magnetic field enhances the small peak around $\mathbf{q}=\mathbf{Q}$ due to the Andreev bound states. Contrastingly, at $\omega=\omega_{res}$, the magnetic field suppresses the resonance at $\mathbf{q}=\mathbf{Q}$ but enhances the intensity from $\mathbf{q}_{x}/\pi=0.7$ to $0.9$.

The behaviors of $\chi^{''}(\mathbf{Q},\omega)$ shown in Figs. \ref{fig1}(a) and \ref{fig2}(a) can be explained as follows: In the SC state, the resonance energy is approximately at
$\omega_{res}=|\Delta(\mathbf{k})|+|\Delta(\mathbf{k}+\mathbf{Q})|$, where $\mathbf{k}$ and $\mathbf{k}+\mathbf{Q}$ are the momenta on the hole- and electron-pockets, respectively. Since the average SC OP in the mixed state is reduced to $0.95\Delta_{0}$, thus the resonance shifts to a lower energy. Furthermore the application of the magnetic field leads to the emergence of in-gap Andreev bound states which induces scattering in the spin gap, thus the spectral weight is transferred from the resonance to that in the spin gap, leading to a weaker and broadened resonance.

On the other hand, to understand the different features of $\Delta\chi^{''}(\mathbf{q},\omega_{res})$ in Figs. \ref{fig1}(d) and \ref{fig2}(d), we study the properties of $\Delta f(i-j,\omega_{res})$ defined as $\Delta f(i-j,\omega_{res})=f(i-j,\omega_{res})|_{mixed-state}-f(i-j,\omega_{res})|_{SC-state}$. For the $s_{\pm}$ pairing, the mixed-state effect influences only the short-range spin-spin correlations [Fig. \ref{fig3}(a)] and the spectrum in Fig. \ref{fig1}(d) can be reproduced by considering only $\Delta f(i-j,\omega_{res})$ in the range $|i_{x}-j_{x}|,|i_{y}-j_{y}|\leq4$. For the $s$-wave case, in contrast, the mixed-state effect extends to much larger spatial areas and all the points in Fig. \ref{fig3}(b) have to be included when performing the Fourier transform in order to match the spectrum in Fig. \ref{fig2}(d). Thus if the real-space spin-spin correlation function can be measured experimentally, Fig. \ref{fig3} can be used as another method to differentiate the $s_{\pm}$ and $s$-wave pairings.

\emph{Summary}.---In summary, we have systematically studied the mixed-state effect on the low-energy spin dynamics in optimally-doped iron pnictide superconductors, for the $s_{\pm}$ and $s$-wave symmetries, respectively. The spin excitation spectra at $\mathbf{q}=\mathbf{Q}$ are similar in these two cases, no matter it is in the normal, SC or mixed state and the evolution of the resonance with the magnetic field are both consistent with the experimental observation, that is, at zero field, a spin gap and a resonance show up in the SC state while in the mixed state, due to the existence of Andreev bound states and the suppressed SC OP, the spin gap is filled up and the resonance shifts to a lower energy and becomes broad, together with a reduced intensity. However, the field-induced intensity change at the resonance energy shows different behaviors between the two symmetries. For the $s_{\pm}$ pairing, the intensity of the spin excitation is suppressed around $\mathbf{q}=\mathbf{Q}$ in the mixed state while for the the $s-$wave case, there are intensity-gain areas in the vicinity of $\mathbf{q}=\mathbf{Q}$. Furthermore, the mixed-state effect affects only the short-range spin-spin correlations in the $s_{\pm}$ case while it extends to much larger spatial areas in the $s-$wave one. We thus propose to measure the field-induced intensity change in momentum space and if possible, in real space, to distinguish the $s_{\pm}$ and $s$-wave symmetries.

This work was supported by NSFC (Grants No. 11204138 and No. 11175087), NSF of Jiangsu Province of China (Grant No. BK2012450), Program of Natural Science Research of Jiangsu Higher Education Institutions of China (Grant No. 12KJB140009), SRFDP (Grant No. 20123207120005), China Postdoctoral Science Foundation (Grant No. 2012M511297), the Texas Center for Superconductivity and the Robert A. Welch Foundation under Grant No. E-1146. The numerical calculations in this paper have been done on the IBM Blade cluster system in the High Performance Computing Center (HPCC) of Nanjing University.

\end{document}